\documentclass{PoS}

\title{The nature of X(3900) and recognition of open charm effects}

\ShortTitle{The nature of X(3900) and recognition of open charm
effects}

\author{\speaker{Qian Wang}\\
Institute of High Energy Physics, Chinese Academy of Sciences, Beijing 100049, China\\
        E-mail: \email{wangqian@ihep.ac.cn}}

\author{Gang Li\\
        Department of Physics, Qufu Normal University, Qufu, 273165, China\\
        E-mail: \email{ligang0314@tom.com}}

\author{Xiao Hai Liu\\
Institute of High Energy Physics, Chinese Academy of Sciences, Beijing 100049, China\\
        E-mail: \email{xhliu@ihep.ac.cn}}

\author{Qiang Zhao\\
Institute of High Energy Physics, Chinese Academy of Sciences, Beijing 100049, China\\
and  Theoretical Physics Center for Science Facilities, CAS, Beijing
100049, China\\
        E-mail: \email{zhaoq@ihep.ac.cn}}

\abstract{We identify open charm effects in a direct production
process $e^+e^-\to J/\psi \pi^0$. A unique feature of this process
is that the $D\bar{D}^*+c.c.$ threshold is located at a relatively
isolated energy region, i.e. $\sim 3.876 \ \mathrm{GeV}$, which is
far away from the well-established charmonia $\psi(3770)$ and
$\psi(4040)$. Therefore, the cross section line-shape of this
reaction provides an opportunity for singling out the open charm
effects. A model-independent narrow enhancement between the
thresholds of $D^0\bar{D}^{*0}+c.c.$ and $D^+D^{*-}+c.c.$ is
predicted. This study can also help understand the $X(3900)$
enhancement recently observed by the Belle and BaBar Collaboration
in $e^+e^-\to D\bar{D}+c.c.$ We also show that the open charm
effects play a crucial role for our understanding of the
long-standing ``$\rho\pi$ puzzle".}

\FullConference{Sixth International Conference on Quarks and Nuclear Physics\\
         April 16-20, 2012\\
         Ecole Polytechnique, Palaiseau,  Paris}

\begin{document}

\section{Introduction}\label{sec-introduction}

The study of hadron spectroscopy covers a broad energy region from
low to high energies, and provides important information about the
dynamics of the strong interaction. The experimental progress during
the past years has brought a lot of surprises to our community.
Taking the charmonium sector as an example, a number of new
resonance-like signals have been observed by the
B-factories~\cite{Olsen:2009ys}. These observations have aroused
great interests from both theory and experiment in understanding
their nature and searching for signals for exotic hadrons (e.g. see
Refs.~\cite{Voloshin:2007dx,Eichten:2007qx,Brambilla:2010cs,Drenska:2010kg}
for a recent review on these issues).

Although various theoretical prescriptions, such as hybrid
charmonium, tetraquark, baryonium, and hadronic molecule, have been
proposed in order to understand the underlying dynamics for the
production and decay of these new charmonium-like ``resonances",  an
obvious feature with those signals is that most of them are close to
open charmed meson thresholds. A typical example is the $X(3872)$
which is located between the $D^0\bar{D}^{*0}$ and $D^+D^{*-}$
thresholds. As a result, a molecular prescription has been broadly
investigated in the literature. It makes the non-perturbative
mechanisms arising from the open charm thresholds an attractive
solution for some of those charmonium-like states.

The Belle Collaboration~\cite{Pakhlova:2008zza} recently observe an
enhancement $X(3900)$ in ISR $e^+ e^-\to D\bar{D}+c.c.$ process. The
interesting feature about this enhancement is that it is directly
produced in the $e^+e^-$ annihilation, thus, its quantum number
should be $J^{PC}=1^{--}$. Meanwhile, one notices that below and
above the $X(3900)$ there are two well established charmonium
states, $\psi(3770)$ and $\psi(4040)$, which can be consistently
accommodated into the charmonium spectrum as $\psi(1D)$ and
$\psi(3S)$ states, respectively. In another word, the $X(3900)$
enhancement is located in a mass region where the quark model does
not have a corresponding $c\bar{c}$ vector state.



Although such an enhancement was conjectured to be caused by the
$D\bar{D}^*+c.c.$ open charm effects, other possibilities seem not
to be eliminated. In Ref.~\cite{Zhang:2009gy}, the $D\bar{D^*}+c.c.$
open charm effects are investigated and the results seem to support
such an explanation without introducing any exotic components.
However, it is not obvious to conclude such a scenario since it is
also shown in Ref.~\cite{Zhang:2009gy} that the enhancement can also
be fitted by a Breit-Wigner structure. In order to clarify the
nature of the $X(3900)$, one should investigate other possible
reflections of such a mechanism. In this proceeding, we present our
recent works on identifying an open charm effect in $e^+ e^-$
annihilations~\cite{Wang:2011yh} and $e^+e^-\to
J/\psi\eta,~J/\psi\pi^0$ and $\phi\eta_c$, where a significant
model-independent enhancement at about $3.876~\mathrm{GeV}$ is
predicted in the process $e^+e^-\to ~J/\psi\pi^0$ as a direct
evidence for the $D\bar{D}^*+c.c.$ open charm effects.

As a dynamical mechanism, the open charm effects can explain the
discrepancies between the experimental data and pQCD predictions
for, e.g. $\psi(3770)$ non-$D\bar{D}$ decay~\cite{Zhang:2009kr} and
helicity selection rule evading
processes~\cite{Wang:2012wj,Liu:2010um,Liu:2009vv}. As a further
evidence for the open charm effects, we also present our recent
studies of the long-standing ``$\rho\pi$ puzzle" in $J/\psi$ and
$\psi^\prime$ decays in this proceeding~\cite{Wang:2012mf}.

\section{The recognition of the open charm effects in the process $e^+e^-\to J/\psi\pi^0$}

\begin{figure}[tb]
\begin{center}
\hspace{-4cm}
\includegraphics[scale=0.5]{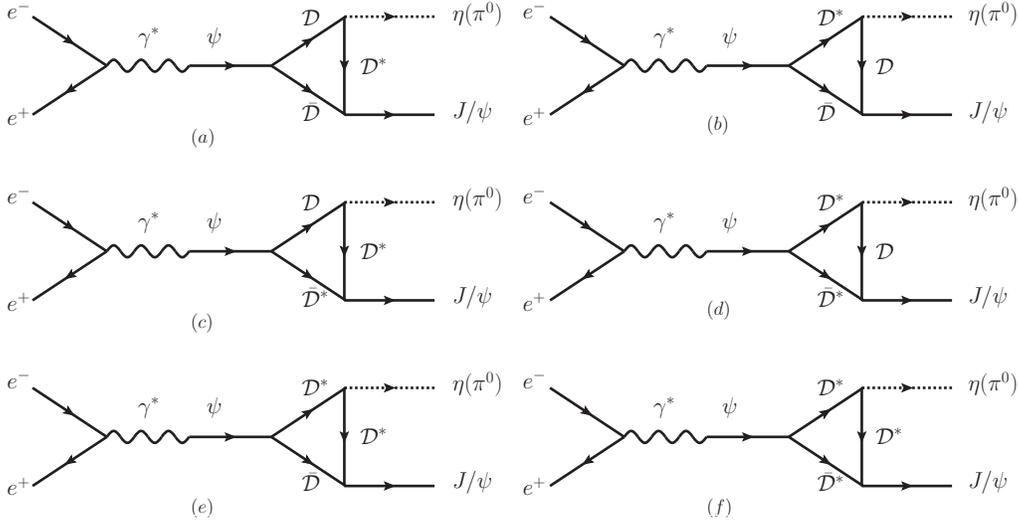}
\vspace{0cm} \caption{Schematic diagrams for $e^+e^-\to
J/\psi\eta(\pi^0)$ via charmed $D$ ($D^*$) meson loops. The diagrams
for the $\phi\eta_c$ mode are similar. }\label{fig:1}
\end{center}
\end{figure}

Based on the effective Lagrangians~\cite{Liu:2010um,Liu:2009vv,
Wang:2010iq,Cheng:2004ru,Casalbuoni:1996pg} and vector meson
dominance (VMD) model, the amplitudes of the diagrams as shown in
Fig.~\ref{fig:1} can be obtained and the details can be found in
Ref.~\cite{Wang:2011yh}. In the energy region that we are interested
in, five vector charmonia, i.e. $J/\psi$, $\psi(3686)$,
$\psi(3770)$, $\psi(4040)$ and $\psi(4160)$, are included as the
dominant pole contributions.

As shown in Fig.~\ref{fig:1}, both neutral and charged $D$ meson
loops can contribute to the $J/\psi\eta$ and $J/\psi\pi^0$
production channels. These two amplitudes have the same sign for the
isospin conserved process, i.e. $J/\psi\eta$ mode, but opposite sign
for the isospin violating $J/\psi\pi^0$ mode. If isospin symmetry is
respected, i.e. $u$ and $d$ quark have the same mass or the charged
and neutral $D$ mesons approximately have the same mass, these two
$D$ meson loops would canceled each other in the isospin violating
process. If not, the charmed meson loops will then give non-zero
residual contribution as a manifestation of the strong isospin
violation.

\begin{figure}[tb]
\hspace{-2.5cm}
\includegraphics[scale=0.7]{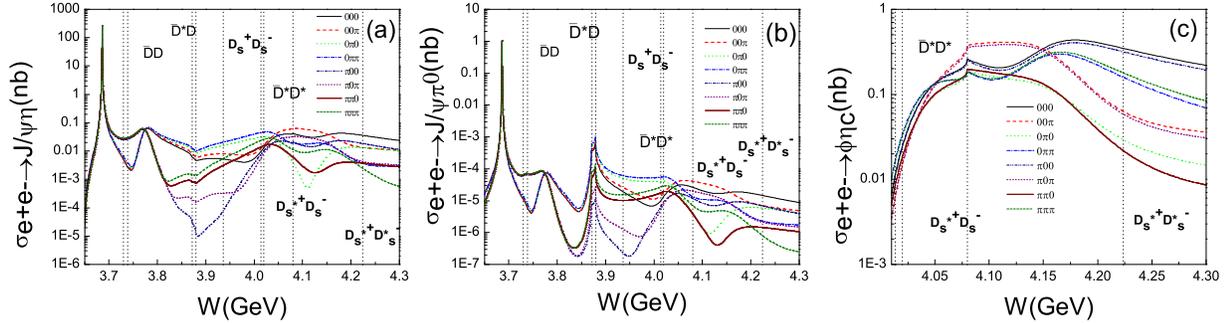}
\vspace{-10cm} \caption{The predicted cross sections for  $e^+e^-\to
J/\psi\eta,~J/\psi\pi^0,~\phi\eta_c$ in terms of the c.m. energy $W$
with the cutoff parameter $\alpha=1.57$ are shown respectively. The
cross sections with different phases, i.e. $(\theta, \beta,\phi)=(0,
0,0), \ (0,0,\pi), \ (0,\pi,0), \ (0,\pi,
 \pi), \ (\pi, 0,0), \ (\pi,0,\pi), \ (\pi,\pi,0), \ (\pi,\pi,
 \pi)$, are presented and denoted by different curves. The vertical lines labels the open charm thresholds. }\label{fig:results}
\end{figure}

Figure~\ref{fig:results}(a) gives the cross section of $e^+e^-\to
J/\psi\eta$ in terms of the c.m. energy. The parameter in the form
factor is obtained by normalizing the cross section at $\psi(3686)$
to the experimental value. These vertical lines are different
charmed meson pair thresholds. Although the neutral and charged
charmed meson loops have a constructive sign, we cannot see a clear
$D\bar{D}^*$ open threshold effect in $e^+e^-\to J/\psi\eta$ due to
large contributions from other processes. Namely, these threshold
effects are submerged by the dominant resonance contributions.

Although the cross sections of these channels are sensitive to the
cut-off parameter $\alpha$, the ratios between them are not. As a
result, with the cut-off parameter $\alpha$ fixed in $e^+e^-\to
J/\psi\eta$, we can predict the cross section of $e^+e^-\to
J/\psi\pi^0$ as shown in Fig.~\ref{fig:results}(b). Since the
charged and neutral $D$ meson loops have opposite signs in the
$J/\psi\pi^0$ production mode, the resonance contributions are
canceled largely. So we can see a much clearer enhancement between
the $D^0\bar{D}^{*0}$ and $D^\pm D^{*\mp}$ thresholds, i.e. $\sim
3.876~\mathrm{GeV}$. Confirmation of such a phenomenon would suggest
that the X(3900) can originate from the $D\bar{D}^*$ open threshold
effects instead of a new charmonium state. Nevertheless, observation
of such an enhancement in $e^+e^-\to J/\psi\pi^0$ should be a direct
evidence for such a non-perturbative dynamic mechanisms.

As a byproduct, we also predict the cross section of $e^+e^-\to
\phi\eta_c$ as shown in Fig.~\ref{fig:results}(c). There is also a
kink structure at the $D_sD_s^*$ threshold. Unfortunately, since it
is near $\psi(4040)$, their interference make this open charm effect
quite insignificant.

\section{A possible explanation of the "$\rho\pi$ puzzle"}

The ``$\rho\pi$ puzzle" in the literature is related to the power
law suppression due to the pQCD helicity selection rule (HSR). For
$J/\psi$ and $\psi'\to VP$, it was shown that these two decays
should be strongly suppressed at leading
twist~\cite{Brodsky:1981kj,Chernyak:1981zz}. Meanwhile, for the
inclusive decays of $J/\psi$ and $\psi'$ into light hadrons via the
$c\bar{c}$ annihilation, it can be related to their leptonic decays
since both processes probe the charmonium wavefunctions at the
origins, i.e.
\begin{small}
\begin{eqnarray}
R\equiv\frac{ BR(\psi^\prime\to hadrons)}{BR(J/\psi \to
hadrons)}\simeq\frac{ BR(\psi^\prime\to e^+e^-)}{BR(J/\psi \to
e^+e^-)}\simeq 0.13 \ ,
\end{eqnarray}
\end{small}
which is the so called ``12\% rule"~\cite{Brodsky:1981kj}. It is
interesting to observe that many exclusive processes respect this
relation quite well, while some processes deviate from this rate
drastically. In particular, the $\rho\pi$  and $K^*\bar{K}+c.c.$
channel have significant deviations from the expectations of both
the pQCD HSR and ``12\% rule". This originates the ``$\rho\pi$
puzzle" in the literature and has attracted a lot of attention from
both theory and experiment.

In our analysis, we put constraints on the electromagnetic (EM)
contribution, short-distance contribution from the $c\bar{c}$
annihilation at the wavefunction origin, and long-distance
contribution from the open charm threshold effects on these two
decays. We show that interferences among these amplitudes, in
particular, the destructive interferences between the short-distance
and long-distance strong amplitudes play a key role to cause the
significant deviations from the pQCD expected ``12\% rule". Since
the mass of $\psi'$ is closer to the open $D\bar{D}$, it would
experience much more significant influences from the long-distance
open charm threshold effects than the $J/\psi$. As a consequence, an
overall suppression of the strong transition amplitudes for
$\psi'\to VP$ can be recognized. Such a phenomenon has also been
realized in some early analyses though the dynamic reason for such a
suppression were not
clarified~\cite{Seiden:1988rr,Suzuki:2001fs,Li:2007ky,Zhao:2006gw}.
In Ref.~\cite{Wang:2012mf} the role played by the open charmed meson
loops are quantified.


Our final results are shown in Table~\ref{tab-results} with
different transitions given explicitly. The charmed meson loop plays
a significant role in $\psi^\prime$ decays, since it closes to the
$D\bar{D}$ threshold. In contrast, it is easy to understand its
negligible influences in $J/\psi$ decays. Meanwhile, the
short-distance parts of these two decays respect the $12\%$ rule.
The destructive interferences between the short-distance and
long-distance part gives the suppression of the strong transition
amplitudes in the $\psi^\prime$ decays, which then become comparable
with the EM contribution.

\begin{table}\footnotesize
\caption{Theoretical results for the branching ratios of $J/\psi \
(\psi')\to VP$ calculated in our model. The experimental data are
from PDG2010~\cite{Nakamura:2010zzi}. }
\begin{tabular}{ccccccc}
  \hline\hline
  $BR(J/\psi\to VP)$ & EM & short-distance & long-distance & strong & total & exp. \\
  \hline
  $\rho\eta$ & $1.81\times10^{-4}$ & $0$ & $2.34\times10^{-12}$ &$2.34\times 10^{-12}$& $1.81\times10^{-4}$ & $(1.93\pm0.23)\times10^{-4}$ \\
  $\rho\eta^\prime$ & $1.37\times10^{-4}$ & $0$ & $2.21\times10^{-12}$ &$2.21\times 10^{-12}$& $1.37\times10^{-4}$ & $(1.05\pm0.18)\times10^{-4}$ \\
  $\omega\pi^0$ & $3.1\times10^{-4}$ & $0$ & $2.38\times10^{-12}$&$2.38\times 10^{-12}$& $3.10\times10^{-4}$ & $(4.5\pm 0.5)\times10^{-4}$ \\
 $\phi\pi^0$ &$9.52\times10^{-7}$ & $0$ & $0$ &0& $9.52\times10^{-7}$ & $<6.4\times10^{-6}$ \\
  $\rho^0\pi^0$ & $4.44\times10^{-5}$ & $4.85\times10^{-3}$ & $2.24\times10^{-7}$ &$4.89\times 10^{-3}$& $5.87\times10^{-3}$ & $(5.6\pm0.7)\times10^{-3}$ \\
  $\rho\pi$ & $1.06\times10^{-4}$ & $1.45\times10^{-2}$ & $6.71\times10^{-7}$ &$1.47\times 10^{-2}$& $1.73\times10^{-2}$ & $(1.69\pm0.15)\times10^{-2}$ \\
  $K^{*+}K^-+c.c.$ & $6.97\times10^{-5}$ & $4.69\times10^{-3}$ & $3.14\times10^{-7}$ &$4.74\times 10^{-3}$& $5.96\times10^{-3}$ & $(5.12\pm0.3)\times10^{-3}$ \\
  $K^{*0}\bar{K}^0+c.c.$ & $1.59\times10^{-4}$ & $4.68\times10^{-3}$ & $3.11\times10^{-7}$ &$4.73\times 10^{-3}$& $3.16\times10^{-3}$ & $(4.39\pm0.31)\times10^{-3}$ \\
  $\omega\eta$ & $1.4\times10^{-5}$ & $1.76\times10^{-3}$ & $1.50\times10^{-7}$ &$1.78\times 10^{-3}$& $2.11\times10^{-3}$ & $(1.74\pm0.20)\times10^{-3}$ \\
  $\omega\eta^\prime$ & $1.4\times10^{-5}$ & $9.91\times10^{-5}$ & $5.42\times10^{-8}$ &$1.02\times 10^{-4}$& $1.92\times10^{-4}$ & $(1.82\pm0.21)\times10^{-4}$ \\
  $\phi\eta$ & $2.35\times10^{-5}$ & $6.70\times10^{-4}$ & $3.22\times10^{-8}$ &$6.76\times 10^{-4}$& $9.52\times10^{-4}$ & $(7.5\pm0.8)\times10^{-4}$ \\
  $\phi\eta^\prime$ & $2.10\times10^{-5}$ & $2.07\times10^{-4}$ & $6.45\times10^{-8}$ &$2.12\times10^{-4}$& $9.93\times10^{-5}$ & $(4.0\pm0.7)\times10^{-4}$ \\
  \hline\hline
  $BR(\psi^\prime\to VP)$ \\
  \hline
  $\rho\eta$ & $1.42\times10^{-5}$ & $0$ & $4.13\times10^{-7}$ &$4.13\times 10^{-7}$& $1.94\times10^{-5}$ & $(2.2\pm0.6)\times10^{-5}$ \\
  $\rho\eta^\prime$ & $1.04\times10^{-5}$ & $0$ & $3.89\times10^{-7}$ &$3.89\times 10^{-7}$& $1.48\times10^{-5}$ & $(1.9^{+1.7}_{-1.2})\times10^{-5}$ \\
  $\omega\pi^0$ & $2.98\times10^{-5}$  & $0$ & $4.25\times10^{-7}$ &$4.25\times 10^{-7}$& $3.73\times10^{-5}$ & $(2.1\pm0.6)\times10^{-5}$ \\
  $\phi\pi^0$ & $9.78\times10^{-8}$ & $0$ & $0$&0& $9.78\times10^{-8}$ & $<4.0\times10^{-6}$ \\
  $\rho^0\pi^0$ & $4.36\times10^{-6}$ & $5.81\times10^{-4}$ & $7.85\times10^{-4}$ &$2.12\times 10^{-5}$& $9.72\times10^{-6}$ & *** \\
  $\rho\pi$ & $1.02\times10^{-5}$ & $1.74\times10^{-3}$ & $2.36\times10^{-3}$ &$6.36\times 10^{-5}$& $3.20\times10^{-5}$ & $(3.2\pm1.2)\times10^{-5}$ \\
  $K^{*+}K^-+c.c.$ & $7.03\times10^{-6}$ &$9.81\times10^{-4}$ & $1.33\times10^{-3}$ &$3.64\times 10^{-5}$& $1.70\times10^{-5}$ & $(1.7^{+0.8}_{-0.7})\times10^{-5}$ \\
  $K^{*0}\bar{K}^0+c.c.$ & $1.61\times10^{-5}$ & $9.81\times10^{-4}$ & $1.39\times10^{-3}$ &$4.61\times 10^{-5}$& $1.09\times10^{-4}$ & $(1.09\pm0.20)\times10^{-4}$ \\
  $\omega\eta$ & $1.10\times10^{-6}$ & $3.24\times10^{-4}$ & $5.57\times10^{-4}$ &$3.52\times 10^{-5}$& $2.48\times10^{-5}$ & $<1.1\times10^{-5}$ \\
  $\omega\eta^\prime$ & $1.12\times10^{-6}$ & $6.23\times10^{-5}$ & $2.31\times10^{-4}$ &$5.43\times 10^{-5}$& $4.01\times10^{-5}$ & $(3.2^{+2.5}_{-2.1})\times10^{-5}$ \\
  $\phi\eta$ & $2.26\times10^{-6}$  & $1.55\times10^{-4}$ & $1.73\times10^{-4}$ &$1.92\times 10^{-6}$& $2.25\times10^{-6}$ & $(2.8^{+1.0}_{-0.8})\times10^{-5}$ \\
  $\phi\eta^\prime$ & $2.22\times10^{-6}$ & $1.85\times10^{-4}$ & $3.99\times10^{-4}$ &$4.33\times 10^{-5}$& $6.42\times10^{-5}$ & $(3.1\pm1.6)\times10^{-5}$ \\
  \hline\hline
\end{tabular}
\label{tab-results}
\end{table}

\section{Summary}

In summary, we present our recent results on the study of the open
charm effects in $e^+e^-\to J/\psi\eta, ~J/\psi\pi^0,~\phi\eta_c$.
We identify a model-independent enhancement at about
$3.876~\mathrm{GeV}$ in $e^+e^-\to J/\psi\pi^0$ due to the isospin
violating contributions via the open charm effects. Such a
prediction, if confirmed, would suggest that the $X(3900)$ may be
due to the $D\bar{D}^*+c.c.$ open threshold effects instead of a new
charmonium state. Such a mechanism can be directly examined in high
statistics measurement of $e^+e^-\to J/\psi\pi^0$. Confirmation of
this mechanism could be essential for our understanding of the
long-standing ``$\rho\pi$ puzzle".

\section{Acknowledgment}

This work is supported, in part, by National Natural Science
Foundation of China (Grant Nos. 11035006 and 10947007), Chinese
Academy of Sciences (KJCX2-EW-N01), Ministry of Science and
Technology of China (2009CB825200), and the Natural Science
Foundation of Shandong Province (Grant No. ZR2010AM011) and the
Scientific Research Starting Foundation of Qufu Normal University.

\end{document}